\documentclass[12pt,preprint2]{emulateapj}

\usepackage{lineno}
\usepackage{natbib}
\usepackage{amsmath}
\usepackage{graphicx}
\usepackage{wasysym}
\usepackage{txfonts}
\usepackage{xspace}
\usepackage{url}
\usepackage{textcomp}
\usepackage{multirow}
\usepackage{lipsum}

\newcommand{\myemail}{ff@caltech.edu}

\newcommand{\cena}{Cen~A\xspace}

\newcommand{\swift}{\textsl{Swift}\xspace}
\newcommand{\fermi}{\textsl{Fermi}\xspace}

\newcommand{\inte}{\textsl{INTEGRAL}\xspace}
\newcommand{\chandra}{\textsl{Chandra}\xspace}
\newcommand{\xmm}{\textsl{XMM-Newton}\xspace}
\newcommand{\sax}{\textsl{BeppoSAX}\xspace}
\newcommand{\suz}{\textsl{Suzaku}\xspace}

\newcommand{\xte}{\textsl{RXTE}\xspace}

\newcommand{\nustar}{\textsl{NuSTAR}\xspace}
\newcommand{\gro}{\textsl{CGRO}\xspace}

\newcommand{\snr}{S/N\xspace}

\newcommand{\msun}{\ensuremath{\text{M}_{\odot}}\xspace}

\newcommand{\redchi}{\ensuremath{\chi^{2}_\text{red}}\xspace}

\newcommand{\feka}{\ensuremath{\mathrm{Fe}~\mathrm{K}\alpha}\xspace}

\newcommand{\nh}{\ensuremath{{N}_\mathrm{H}}\xspace}

\newcommand{\nhone}{\ensuremath{{N}_{\mathrm{H},1}}\xspace}
\newcommand{\nhtwo}{\ensuremath{{N}_{\mathrm{H},2}}\xspace}

\newcommand{\asec}{\ensuremath{''}\xspace}
\newcommand{\ergcms}{\ensuremath{\mathrm{erg\,cm^{-2}\,s^{-1}}}}

\shorttitle{\nustar and \xmm observations of Cen A }
\shortauthors{F\"urst et al.}

\begin{document}

\title{\nustar and \xmm observations of the hard X-ray spectrum of Centaurus A}

\author{F.~F\"urst\altaffilmark{1} }
\author{C.~M\"uller\altaffilmark{2,3,4} }
\author{K.~K.~Madsen\altaffilmark{1} }
\author{L.~Lanz\altaffilmark{5} }
\author{E.~Rivers\altaffilmark{1} }
\author{M.~Brightman\altaffilmark{1} }
\author{P.~Arevalo\altaffilmark{6}}
\author{M.~Balokovi\'c\altaffilmark{1}}
\author{T.~Beuchert\altaffilmark{4,3} }
\author{S.~E.~Boggs\altaffilmark{7}}
\author{F.~E.~Christensen\altaffilmark{8}}
\author{W.~W.~Craig\altaffilmark{7,9}}
\author{T.~Dauser\altaffilmark{4}}
\author{D.~Farrah\altaffilmark{10}}
\author{C.~Graefe\altaffilmark{4,3}}
\author{C.~J.~Hailey\altaffilmark{11}}
\author{F.~A.~Harrison\altaffilmark{1}}
\author{M.~Kadler\altaffilmark{3}}
\author{A.~King\altaffilmark{12}}
\author{F.~Krau\ss\altaffilmark{4,3}}
\author{G.~Madejski\altaffilmark{12}}
\author{G.~Matt\altaffilmark{13}}
\author{A.~Marinucci\altaffilmark{13}}
\author{A.~Markowitz\altaffilmark{14}}
\author{P.~Ogle\altaffilmark{5}}
\author{R.~Ojha\altaffilmark{15,16,17}}
\author{R.~Rothschild\altaffilmark{14}}
\author{D.~Stern\altaffilmark{18}}
\author{D.~J.~Walton\altaffilmark{18,1}}
\author{J.~Wilms\altaffilmark{4}}
\author{W.~Zhang\altaffilmark{15}}

\altaffiltext{1}{Cahill Center for Astronomy and Astrophysics, California Institute of Technology, Pasadena, CA 91125, USA (\myemail)}
\altaffiltext{2}{Department of Astrophysics/IMAPP, Radboud University Nijmegen, 6500 GL, Nijmegen, The Netherlands}
\altaffiltext{3}{Lehrstuhl f\"ur Astronomie, Universit\"at W\"urzburg, 97074 W\"urzburg, Germany}
\altaffiltext{4}{Dr. Karl-Remeis-Sternwarte and ECAP,  96049 Bamberg, Germany}
\altaffiltext{5}{Infrared Processing and Analysis Center, California Institute of Technology, Pasadena, CA 91125, USA}
\altaffiltext{6}{Instituto de F\'isica y Astronom\'ia, Facultad de Ciencias, Universidad
de Valpara\'iso,  Valpara\'iso, Chile}
\altaffiltext{7}{Space Sciences Laboratory, University of California, Berkeley, CA 94720, USA}
\altaffiltext{8}{DTU Space, National Space Institute, Technical University of Denmark, 2800 Lyngby, Denmark}
\altaffiltext{9}{Lawrence Livermore National Laboratory, Livermore, CA 94550, USA}
\altaffiltext{10}{Department of Physics, Virginia Tech, Blacksburg, VA 24061, USA}
\altaffiltext{11}{Columbia Astrophysics Laboratory, Columbia University, New York, NY 10027, USA}
\altaffiltext{12}{Kavli Institute for Particle Astrophysics and Cosmology,
Stanford University, Menlo Park,
CA 94025, USA}
\altaffiltext{13}{Dipartimento di Matematica e Fisica, Universit\`a degli Studi
Roma Tre, 00146 Roma, Italy}
\altaffiltext{14}{University of California, San Diego, CASS,  La Jolla, CA 92093, USA}
\altaffiltext{15}{NASA Goddard Space Flight Center, Greenbelt, MD 20771, USA}
\altaffiltext{16}{University of Maryland, Baltimore County, Baltimore,MD 21250, USA}
\altaffiltext{17}{The Catholic University of America, Washington, DC 20064}
\altaffiltext{18}{Jet Propulsion Laboratory, California Institute of Technology, Pasadena, CA 91109, USA}

\begin{abstract}
We present simultaneous \xmm and \nustar observations spanning 3--78\,keV of the nearest radio galaxy, Centaurus~A (Cen~A). The accretion geometry around the central engine in \cena is still debated, and we investigate possible configurations using detailed  X-ray spectral modeling.
\nustar imaged the central region of \cena with subarcminute resolution at X-ray energies above 10\,keV for the first time, but finds no evidence for an extended source or other off-nuclear point-sources. 
The \xmm and \nustar spectra agree well and can be described with an absorbed power-law with a photon index $\Gamma=1.815\pm0.005$ and a fluorescent \feka line in good agreement with literature values. 
The spectrum does not require a high-energy exponential rollover, with a constraint of $E_\text{fold} > 1$\,MeV.
A thermal Comptonization continuum describes the data well, with parameters that agree with values measured by \inte, in particular an electron temperature $kT_e$ between $\approx$100--300\,keV, seed photon input temperatures between 5--50\,eV. We do not find evidence for reflection or a broad iron line and put stringent upper limits of $R<0.01$ on the reflection fraction and accretion disk illumination.
We use archival \chandra data to estimate the contribution from  diffuse emission, extra-nuclear point-sources, and the outer X-ray jet to the observed \nustar and \xmm X-ray spectra and 
find the contribution to be negligible. 
We discuss different scenarios for the physical origin of the observed hard X-ray spectrum, and conclude that the inner disk is replaced by an advection-dominated accretion flow  or that the X-rays are dominated by synchrotron self-Compton emission from the inner regions of the radio jet or a combination thereof. 
\end{abstract}


\section{Introduction}
At a distance of 3.8\,Mpc \citep{Harris2010}, Centaurus~A (Cen~A,
PKS~1322$-$428, NGC~5128) is the closest active galaxy exhibiting
powerful jets.  It hosts a
supermassive black hole with a mass of $M \sim 5\times 10^7 M_\odot$, as estimated from dynamical modeling of the gas disk surrounding the black hole
\citep{Neumayer2007}. \cena is bright across the electromagnetic spectrum and among the first identified extragalactic X-ray sources \citep{bowyer70a}. In recent years, it has been 
detected up to $\gamma$-ray energies by \textsl{Fermi}/LAT
\citep{Abdo2010CenAlobes,fermicena10a} and
H.E.S.S. \citep{Aharonian2009}. Due to its proximity, it is an ideal
laboratory to study the physics of active galactic nuclei (AGN)  including jet-launching mechanisms and coronal geometry 
\citep[see][for an extensive review]{Israel1998a}. 

\cena shows a complex structure, revealed at
different wavelengths. Optical observations reveal a prominent dust
band across the giant elliptical host galaxy NGC~5128, possibly indicating
a merger event \citep[e.g.,][]{Israel1998a}.
Powerful radio lobes are seen, extending almost perpendicular to this dust lane  out to a
projected size of 10$^\circ$
on the sky (corresponding to about 600\,kpc at the distance of \cena). It is classified as a proto-typical
Fanaroff-Riley type~I radio galaxy \citep[FR~I,][]{Fanaroff1974}.  

Jets are observed and resolved from the radio up to X-ray
energies. High-resolution radio observations probe the jet
in detail from
sub-parsec to kilo-parsec scales
\citep[e.g.,][]{Mueller2014,Feain2011,Hardcastle2003,kraft02a}.
The X-ray jet, extending about $2'$, shows a knotty substructure with spectral
steepening to the jet edges \citep{Hardcastle2003,Worrall2008a}. It is resolved down to about 50\,ly from the core, at which point it becomes invisible over the core emission even in \chandra.

The soft X-ray (0.1--7\,keV) morphology of \cena shows a very bright AGN, a fainter jet, and surrounding diffuse emission. The diffuse emission originates from the hot interstellar medium (ISM), which is measurable as a soft thermal component in the X-ray spectrum, as well as from off-nuclear point sources, mostly low-mass X-ray binaries \citep{kraft03a}.
Accretion takes place at very low Eddington fractions \citep[$<0.2\%$,][]{evans04a}, allowing a classification as a low-luminosity radio galaxy.

The broad-band X-ray spectrum of \cena is  complex, consisting of several
emission components, in particular a soft thermal plasma  at low energies (0.1--2\,keV), a power-law continuum, and strong absorption. Their origin is still unclear, including whether
the hard X-ray spectrum solely originates from Comptonization in a thermal corona close to the core or  also has a jet
synchrotron self-Compton (SSC) component from the inner jet, unresolved in X-rays \citep[e.g.,][]{markowitz07a,fermicena10a,fukazawa11a,Mueller2014}. The location and structure
of the absorbing material is also still uncertain, and partial-covering
models have been discussed \citep[e.g.,][]{evans04a,markowitz07a,fukazawa11a}. Further, \cena shows strong
$N_\mathrm{H}$ variations with time indicating a clumpy torus
\citep{Markowitz2014,rivers11a,rothschild11a}.

The hard power-law continuum ($\sim$3--100\,keV) can be well described by a power law with a spectral
index of $\Gamma\sim1.8$ with an average unabsorbed flux of $\mathcal{F}_\mathrm{20-100\,keV}\approx 6\times
  10^{-10}\,\ergcms$, attenuated by strong absorption (typical \nh values $>10^{23}$\,cm$^{-2}$) at energies below 10\,keV \citep[see, e.g.][and references therein]{ mushotzky78a,baity81a, rothschild11a, beckmann11a}.
On top of the continuum a strong \feka line is
present, with an equivalent width of typically $\sim80$\,eV
\citep{markowitz07a,fukazawa11a}. 

 Fluorescent \feka lines are
often a tell-tale sign of reflection off dense material in AGN and are
commonly observed \citep[e.g.,][]{singh11a}. However, reflection
off the accretion disk or optically thick torus  also leads to the production of a Compton hump
between 10--30\,keV \citep{ross05a}, the existence of which is debated in \cena \citep[and references therein]{rivers11a, fukazawa11a}.
Furthermore, the observed \feka line in \cena is always narrow, ruling out an origin close to the central black hole.

Based on \suz data, \citet{fukazawa11a} report the detection of reflection,
i.e., a Compton hump, when introducing a second  power-law component ($\Gamma < 1.6$) to describe the continuum.
Using \chandra and \inte/SPI data \citet{burke14a} come to a similar conclusion.
However, \citet{beckmann11a}, using all \inte instruments  do not find a significant
reflection component as modeled by \texttt{pexrav} \citep{magdziarz95a} and put a 3$\sigma$ upper
limit of $R<0.28$.  Here, $R$ is the reflection fraction which is defined as 1 for reflection off an infinite disk, i.e., a reflector covering $2\pi$ of the sky as seen from the primary X-ray source.
Applying a physically-motivated Comtponziation model \citep[\texttt{compPS},][]{poutanen96a}, \citet{beckmann11a} find weak
evidence for reflection with $R=0.12^{+0.08}_{-0.10}$, which is still
consistent with no reflection at the $1.6\,\sigma$ level. 

\citet{rothschild11a} studied over 12 years of \textsl{Rossi X-ray Timing Explorer} (\xte) data and find a
very stable photon-index $\Gamma=1.822\pm0.004$, despite significant
variation in the X-ray flux, and no evidence for reflection. They argue that the line is likely produced in
a Compton-thin torus, thereby not producing a measurable
Compton hump. These findings are confirmed by \citet{rivers11a}, who put an upper limit of $R<0.005$ on the reflection fraction using \xte.

\citet{evans04a} use \chandra and \xmm data to study the soft X-ray
spectrum of \cena in detail. They use heavily piled-up \xmm data of
two different observations taken in 2001 and 2002 from which they
excised the inner 20\asec to reduce pile-up. Additionally they add the
diffuse emission as measured by \chandra to the \xmm background to
obtain a clear measurement of the core spectrum. 
They find that for an accurate description of the \xmm spectrum two
absorbed power-law components are required, with the primary one
having a photon-index of $\Gamma_1 = 1.74^{+0.11}_{-0.09}$ and an
absorption column of $\nhone =
\left(1.19\pm0.13\right)\times10^{23}$\,cm$^{-2}$. For the second
power-law they fixed the photon-index $\Gamma_2 = 2$ and measured an
absorption column of $\nhtwo =
\left(3.6^{+2.2}_{-2.3}\right)\times10^{22}$\,cm$^{-2}$.

Here, we present simultaneous \textsl{Nuclear Spectroscopic Telescope Array} \citep[\nustar,][]{harrison13a} and \xmm \citep{jansen01a} observations taken in August 2013 to study the AGN core (see Table~\ref{tab:obsdates}). \nustar is ideally suited to study reflection spectra in AGN since it covers the \feka line region and the Compton hump with one instrument.
 This allows us to investigate the accretion geometry and the physics of the central engine through detailed spectral modeling. 
We also use archival quasi-simultaneous \chandra data to study possible contamination from the diffuse and point-source emission.

The remainder of the paper is organized as follows: in Section~\ref{sec:obs} we give an overview over the data used and data reduction procedures. In Section~\ref{sec:image} we present X-ray images and search for extended emission at high energies. In Section~\ref{sec:numodel} we describe the spectral modeling, including the contribution from the diffuse emission. We discuss our findings in Section~\ref{sec:disc} and summarize the results in Section~\ref{sec:summ}.
We adopt a redshift of $z=0.0018$ throughout the paper and give errors at the 90\% confidence level for one parameter of interest unless otherwise noted. Data analysis was performed with  the Interactive Spectral Interpretation System v1.6.2-30 \citep[ISIS;][]{houck00a}.

\section{Observations and data reduction}
\label{sec:obs}

\begin{deluxetable}{lccc}
\centering
\tablecolumns{8}
\tabletypesize{\scriptsize}
\tablecaption{Observation log showing the observation number for each observatory as well as the exposure time for each instrument. \label{tab:obsdates}}
\tablehead{ \colhead{ObsID}  &  \colhead{MJD range} & Instrument & \colhead{exp. time [ks]}}
\startdata 
\cutinhead{\nustar}
60001081002 & 56510.54--56511.67 & FPMA & 51.26 \\
& & FPMB & 51.35 \\
\cutinhead{\xmm}
0724060601 & 56511.53--56511.66 & EPIC-pn & 7.29 \\
& & MOS\,1 & 10.50 \\
& & MOS\,2 & 10.49 \\
\cutinhead{\chandra (see Appendix)}
7797, 7798, 7799, 7800 & 54181.37--54207.63 & ACIS-I & 373.35 \\
15295 & 56535.91--56536.01 & ACIS-I & 5.35 \\
\enddata
\end{deluxetable}

\subsection{\nustar}
 \label{susec:nustar}
\nustar consists of two independent grazing incidence telescopes, focusing X-rays between 3--78\,keV on corresponding focal planes consisting of cadmium zinc telluride (CZT) pixel detectors. \nustar provides unprecedented sensitivity and high spectral resolution at energies above 10\,keV, ideally suited to study the Compton reflection hump. The two focal planes are referred to as focal plane modules (FPM) A and B. \nustar data were extracted using the standard \texttt{NUSTARDAS} v1.3.1 software. Source spectra were taken from a $100''$ radius region center on the J2000 coordinates. The background was extracted as far away from the source as possible, from a $120''$ radius region. This approach induces small systematic uncertainties in the background, as the background is known to change over the field of view \citep{wik14a}. However, \cena is over a factor $\sim$10 brighter than the background even at the highest energies, so that these uncertainties are negligible. \nustar data were binned to a signal-to-noise ratio (\snr) of 20 in the relevant energy range of 3--78\,keV within ISIS.

The average count-rate during the observations was $\approx$18.5\,cts\,s$^{-1}$ per module. Only very slight variability was evident, with the count-rate declining by about 5\% over the observation. No changes in hardness were visible, so we use the time-averaged spectrum for the remainder of this paper. 

\subsection{\xmm}
\xmm observed \cena as part of the Tracking Active Galactic Nuclei with Austral Milliarcsecond
Interferometry program  (TANAMI), an on-going multi-wavelength, multi-year monitoring program of southern AGN \citep{ojha10a, Mueller2014}.
We reduced the \xmm data using the standard scientific analysis software (SAS) version \verb+xmmsas_20141104_1833-14.0.0+. The EPIC-pn camera \citep{strueder01a} was operated in a small window mode to eleviate pile-up while the MOS cameras \citep{turner01a} were operated in the full frame mode to obtain a measurement of the diffuse and jet components. A detailed analysis of the \xmm data will be presented in a forthcoming publication (M\"uller et al., in prep.). Here we concentrate on the energy range $>3$\,keV for a direct comparison with the \nustar data and to avoid contamination from the soft X-ray emission from the thermal extended plasma and the off-nuclear point-sources.

Even though EPIC-pn was operated in the small window mode, the count-rate of $\approx 30$\,cts\,s$^{-1}$ is enough to cause pile-up (see the \xmm users' handbook issue 2.13\footnote{\url{http://xmm.esac.esa.int/external/xmm\_user\_support/documentation/ uhb/index.html}}). We therefore carefully analyzed extraction regions with different annuli and compared spectral shapes and the results from \texttt{epatplot}. We found that only negligible fractions of pile-up remain for an inner radius of 10\asec. We set the outer radius to 40\asec, the largest radius possible with the region fully on the chip, as the source was located close to the north-east border of the chip. We rebinned the pn data to a \snr of 15 between 3--10\,keV.

Having been operated in full window mode, MOS\,1 and 2 were more significantly piled-up, and we excluded the inner 20\asec to remove most pile-up effets. We set the outer radius to 100\asec to be comparable to the \nustar extraction region and rebinned the spectra to a \snr of 11.5 between 3--9\,keV to retain sufficient spectral resolution for line spectroscopy despite the lower effective area compared to pn. Within that annulus, no other point source is visible.
A more detailed study of the jet spectrum including \chandra will be presented in a forthcoming publication (Graefe et al., in prep.).

All annuli were centered on the J2000 coordinates of \cena.
The \xmm data were taken contemporaneously to \nustar, overlapping in the last part of the longer \nustar observation. The complete observation log is given in Table~\ref{tab:obsdates}.

\section{Imaging}
\label{sec:image}

\begin{figure*}
\centering
\includegraphics[width=0.95\textwidth]{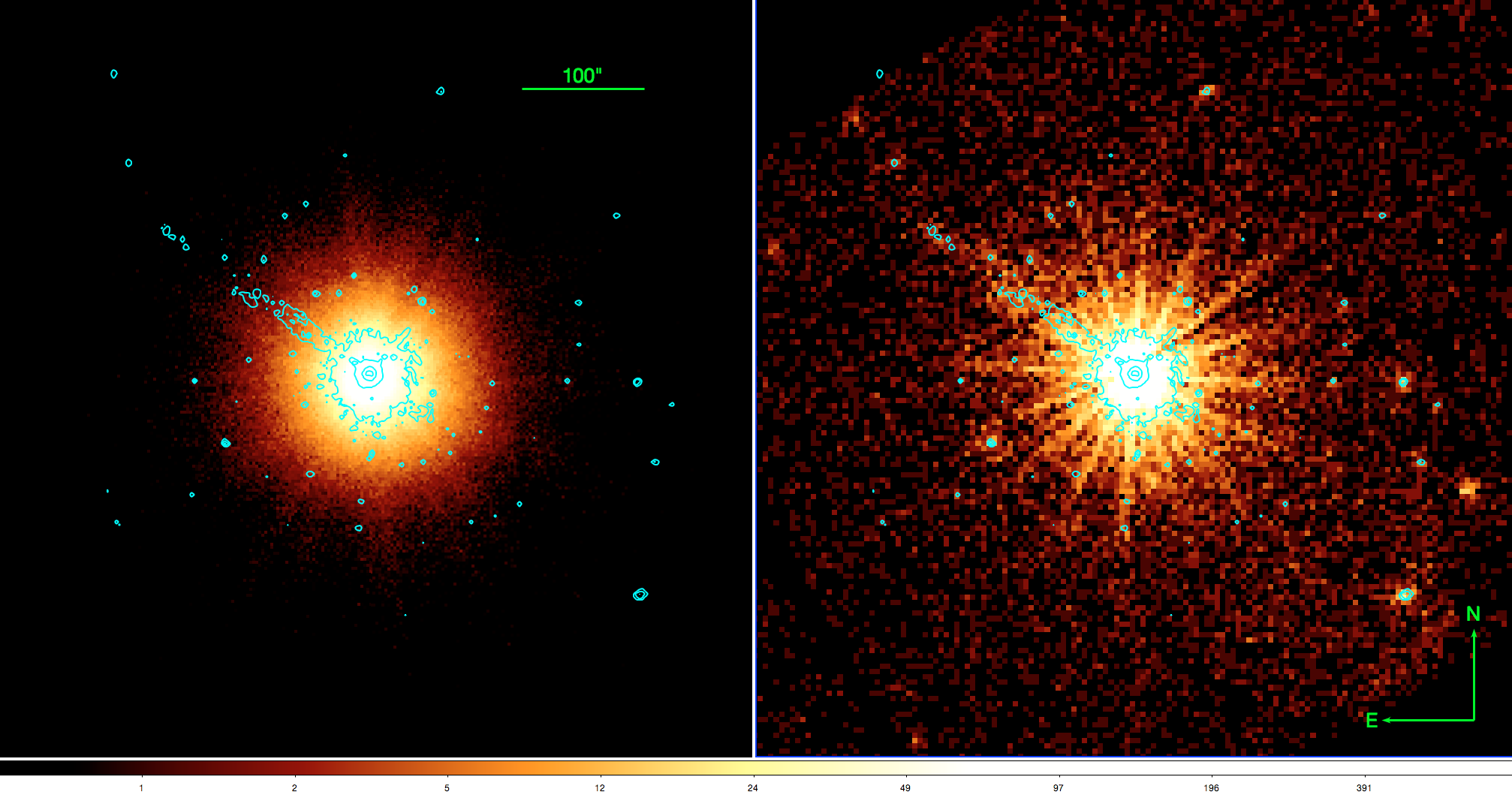}
\caption{\nustar FPMA (left) and \xmm MOS\,1 (right) images of the \cena core. North is up, East is to the left. Superimposed in cyan are the \chandra contours. The jet extends to the north-east and is faintly detected in the MOS image. 
 }
\label{fig:image}
\end{figure*}

We show the \nustar image in the 3--78\,keV energy band in the left panel of Figure~\ref{fig:image}, which is consistent with a point source. Even after careful deconvolution of the image, we find no evidence for a deviation from a point source. In particular, the outer jet is not visible in the \nustar data. This is mainly due to the broad point-spread function (PSF) of \nustar with a half-power diameter of 60\asec \citep{nustarcalib}. The PSF smears out the very bright core over most of the bright jet emission. When summing up the counts observed by \chandra in knots AX and BX, as described by \citet{kraft00a}, we would expect a count-rate of  $\approx5\times10^{-3}$\,counts\,s$^{-1}$\,module$^{-1}$ in \nustar. However, we measure 0.75\,counts\,s$^{-1}$\,module$^{-1}$ in the jet region, i.e., almost two orders of magnitude larger. The counts in this region are completely dominated by the core emission and the Poissonian noise is of the same order as the expected jet count-rate.
\chandra analysis also indicates that the jet is mainly visible in the soft X-rays (Graefe et al., in prep.), making a detection above $>3$\,keV unlikely. 

The right panel of Figure~\ref{fig:image} shows the MOS\,1 image together with the X-ray contours from \chandra. While the MOS PSF has  a half energy width  of only $13''$, the \cena core is so bright that it contributes significantly to the image out to at least $110''$. The spikes surrounding the core in the image are due to the X-ray optics. \cena's jet extends to the north-east and can be made out in the MOS data. The diffuse emission as observed by \chandra is too weak to contribute visibly to the image. Note also that the off-nuclear point-sources (e.g., in the south-west corner) are not visible in \nustar.

\section{Spectral modeling}
\label{sec:numodel}

We modeled the \nustar FPMA and B and the \xmm EPIC-pn, MOS\,1 and 2 data taken in August 2013 simultaneously with  ISIS, allowing for cross-calibration constants between the instruments ($CC_i$). We give all fluxes relative to FPMA ($CC_\text{FPMA} = 1$). The \nustar and \xmm data show big discrepancies between 3--5\,keV, with \nustar measuring a significantly higher flux than the \xmm instruments. This discrepancy has also been observed in other simultaneous data as well as with \swift/XRT and is at the time of writing being investigated by the \nustar team (Madsen et al., in prep.). We ignore \nustar data below 5\,keV for now, as the \xmm EPIC-pn data provide data with a higher \snr (but see Section~\ref{susec:diff}).
We consequently use \nustar between 5--78\,keV, \xmm pn between 3--10\,keV, and MOS between 3--9\,keV.

\subsection{Point-source emission}
We first fit the data with an absorbed power-law as shown in Fig.~\ref{fig:spec_total}. A prominent \feka line is visible in the residuals (Fig.~\ref{fig:spec_total}b), which can be described with a narrow Gaussian around 6.4\,keV with an equivalent width of $\approx40$\,eV. The Gaussian is narrower than the energy resolution of \xmm and we only find upper limits for its width. 
The absorption is modeled with the \texttt{phabs} model, using abundances by \citet{wilms00a} and cross-sections by \citet{verner96a}. This model gives a good fit (\redchi =1.04 for 1532 dof) with a power-law index $\Gamma= 1.815\pm0.005$. We calculate an unabsorbed 3--50\,keV luminosity of $\approx3.4\times10^{42}$\,erg\,s$^{-1}$.
 All parameters can be found in Table~\ref{tab:par_full}.
 Note that uncertainties are purely statistical and do not take systematic differences between the detectors into account \citep[e.g., the photon-index can vary by $\approx 0.01$ between consecutive observations in \nustar and the line energies have about 15\,eV systematic uncertainties, see][]{nustarcalib}.

To investigate the process responsible for the hard X-ray continuum and estimate the coronal temperature in a thermal Comptonization scenario, we searched for the presence of a exponential rollover at high energies by replacing the power-law with the \texttt{cutoffpl} model in XSPEC. The fit did not improve  and we obtained a lower limit of $E_\text{fold} > 1$\,MeV (see Table~\ref{tab:par_full}). This limit is far above the \nustar energy range and therefore  unreliable. However, as the \texttt{cutoffpl} is only a phenomenological model which shows continuous curvature even far below the folding energy, this result indicates that the 3--78\,keV spectrum of \cena is a pure power-law.

For a more realistic description of a continuum produced by Comptonization, we applied the \texttt{compps} model \citep{poutanen96a}. Following \citet{beckmann11a}, we assume a multi-colored disk with a slab geometry and fit for the Compton-$y$ parameter. 
The disk input temperature  cannot be constrained with our data due to obscuration,  so we in a first approach fix it at $kT_\text{BB}$=10\,eV, appropriate for a black hole mass of $5\times10^7$\,\msun accreting at very low Eddington fractions \citep{makishima00a}.
The \texttt{compps} model also includes a reflection component based on the \texttt{pexrav} model and described by the reflection strength $R$, which we allow to vary. The inclination\footnote{here $i=0^\circ$  corresponds to a face-on view, while $i=90^\circ$  corresponds to an edge-on view}  was set to $i=60^\circ$.  To describe the \feka line, we added a Gaussian component and obtained a very good fit, with $\redchi=1.04$ for 1531 dof. The values obtained for $y=0.402\pm0.016$ and the coronal temperature $kT_e=216^{+19}_{-22}$\,keV agree very well with the results from \citet{beckmann11a}, see Table~\ref{tab:par_full}. We only find an upper limit on the reflection strength at the 90\% confidence level of $R\le0.012$.

We investigated the influence of the disk input temperature on other parameters within a reasonably expectable a range, sampling temperatures between $kT_\texttt{BB}$=5--50\,eV. We find that the plasma temperature to first order decreases with hotter disk temperatures, from $277^{+21}_{-26}$\,keV at 5\,eV to $118^{+13}_{-14}$\,keV at 50\,eV. At higher input temperatures, however,  a secondary minimum evolves at high plasma temperatures around 350\,keV, which becomes statistically preferred above $\sim$60\,eV.  At $kT_\texttt{BB} = 100$\,eV we then measure an electron temperature of $304^{+19}_{-16}$\,keV. We note that a disk temperature above 50\,eV are likely too high for the parameters of \cena's black hole and we therefore do not investigate this solution further.

Using the \texttt{comptt} model  \citep{titarchuk94a}  only gives a  lower limit of $kT_e > 475$\,keV.
The measured value of the electron temperature should be taken with a grain of salt and is strongly influenced by our assumptions.  A full investigations of the systematic uncertainties is, however, beyond the scope of this paper.

\begin{figure}
\centering
\includegraphics[width=0.95\columnwidth]{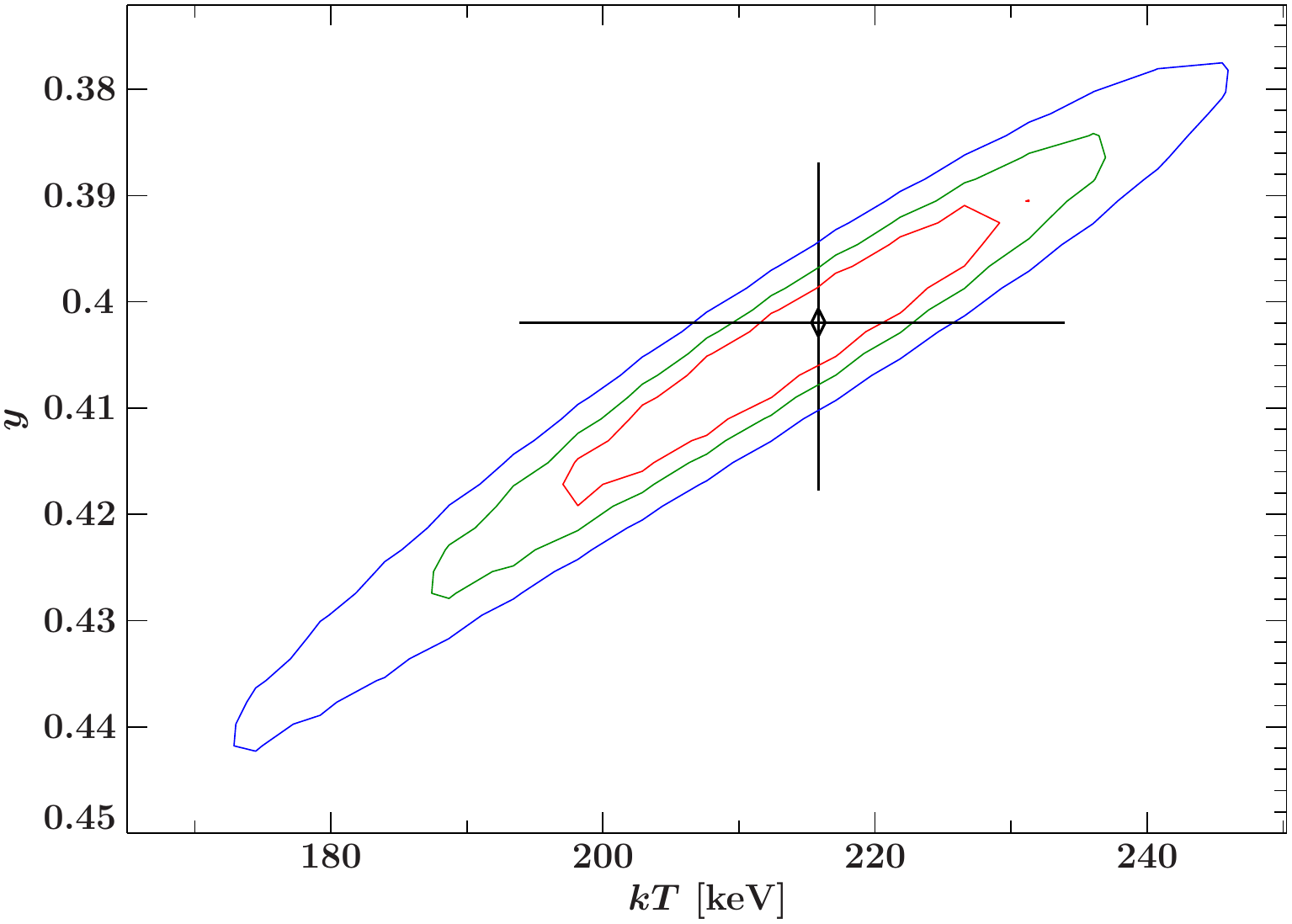}
\caption{Contour map of the electron temperature $kT_e$ versus the Comtpon-$y$ parameter in terms of $\chi^2$. The black cross indicates the best-fit and its 90\% uncertainties. The contours are given at the $1\sigma$, 90\%, and 99\% confidence level (red, green, and blue, respectively).}
\label{fig:cm_kt2y}
\end{figure}

\begin{figure}
\centering
\includegraphics[width=0.95\columnwidth]{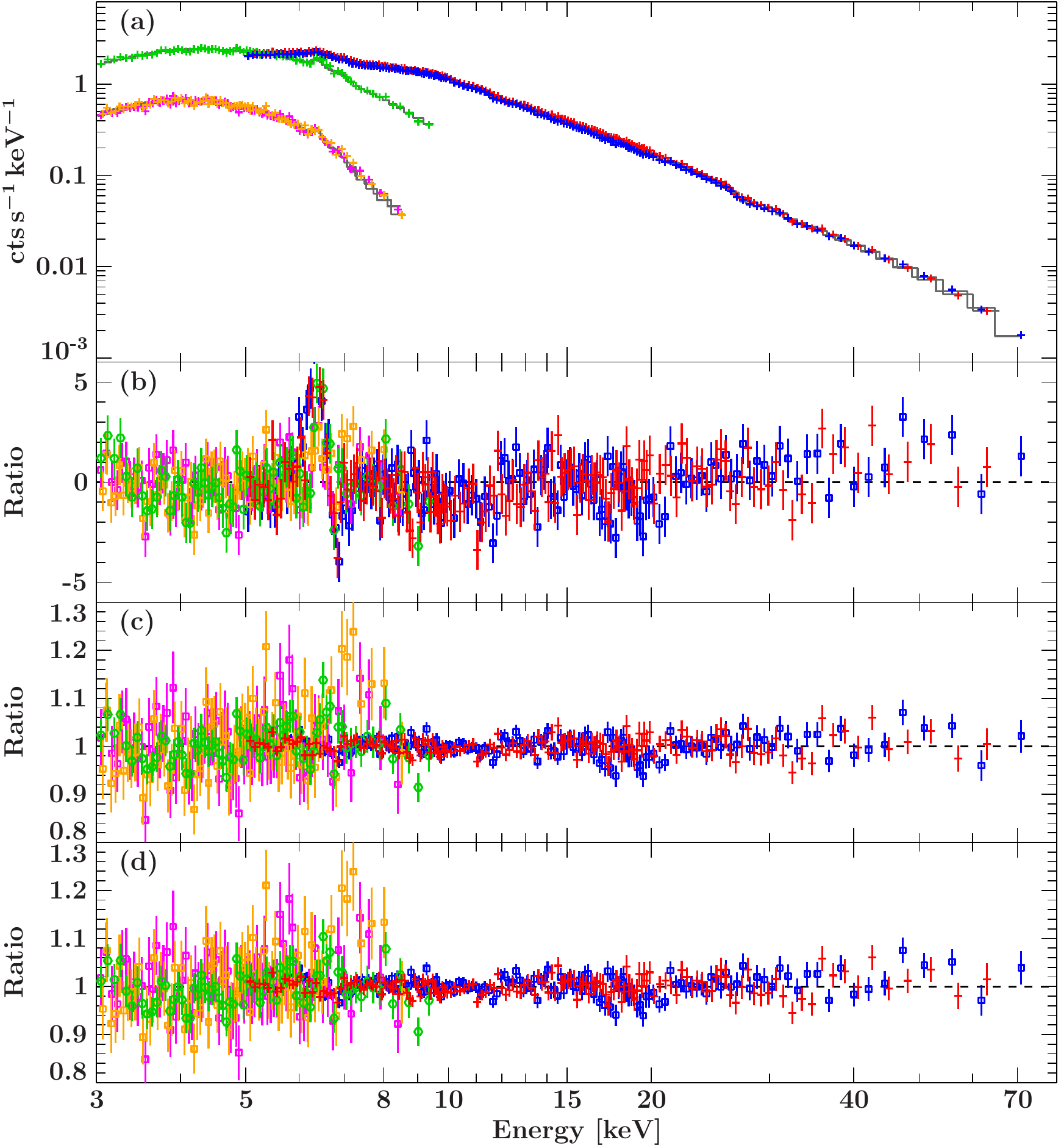}
\caption{\textit{(a)}: \xmm pn (green), MOS\,1 (orange), and MOS\,2 (magenta) as well as \nustar FPMA (red) and FPMB (blue) spectra. The best-fit power-law model with an additional \feka line is superimposed in gray. \textit{(b)} residuals in terms of $\chi$ for the best-fit power-law model without the \feka line (MOS\,1 and 2 show similar residuals but are not shown for clarity). \textit{(c)} residuals in terms of ratio to the best-fit power-law model, including the \feka line. For details see text. \textit{(d)} residuals for the best-fit \texttt{compps} model. Data were rebinned for visual clarity.}
\label{fig:spec_total}
\end{figure}

Despite the fact that electron temperature is above the energy range covered by \nustar, we can constrain $kT_e$ for a given disk temperature, due to the spectral shape and the high \snr of our data. In Figure~\ref{fig:cm_kt2y} we show the $\chi^2$ confidence contours for $kT_e$ versus the Compton-$y$ parameter, assuming $kT_\text{BB} = 10$\,eV. While a clear degeneracy can be seen, both parameters are well constrained. When we directly fit for the optical depth $\tau$ instead of $y$, we find a very similar contour and a best-fit value of  $\tau= 0.240^{+0.041}_{-0.027}$. 


To test if the intrinsic shape of the \texttt{compps} is concealing any weak reflection component, we then modeled the spectrum using only the \texttt{pexrav} model. We note that the \texttt{pexrav} model does not make any assumptions about the geometry of the reflecting medium and is therefore also mostly a phenomenological model to test for the presence of curvature. 
Again, we only obtain an upper limit of $R\le0.010$ on the reflection fraction (Table~\ref{tab:par_full}). The value depends on the assumed inclination (here we used an inclination of $i=60^\circ$, following \citealt{fukazawa11a}) and is even lower for smaller angles ($\approx$0.006 at the model maximum $\cos(i)=0.95$). This limit is similar to the one obtained by \citet{rivers11a} using \xte data ($R<0.005$).

Models that self-consistently describe the reflection spectrum off an optically thick disk, like \texttt{pexmon} \citep{nandra07a}, \texttt{reflionx} \citep{rossandfabian} and \texttt{xillver} \citep{garcia10a} and include line fluorescence and a Compton hump, fail to provide an adequate description of the spectrum within  physically sensible  parameters. These models cannot combine the strength of the iron line with the lack of a Compton hump, indicating that the \feka line does not originate from reflection off Compton thick material.

Finally, we tested physically motivated models for the presence of a toroidal obscuring structure in the nuclear region of \cena. We applied the X-ray spectral models of \citet[BNTorus]{brightman11a} and \citet[MYTorus]{murphy09a}, which were designed specifically for this purpose. The models self-consistently account for photo-electric absorption, fluorescence line emission (most importantly from \feka) and Compton scattering, assuming a toroidal geometry. The MYTorus model assumes an obscuring torus with a circular cross-section and a fixed opening angle 
$\Theta_\text{tor}$ of $60^\circ$, while the BNTorus model assumes a spherical torus where \nh is independent of the inclination (i.e., viewing angle) $i$. The spherical torus is modified by a biconical void with a variable opening angle  $\Theta_\text{tor}$. Furthermore, BNTorus allows for variation of the covering factor of the torus, whereas MYTorus has a fixed covering factor of 0.5. 
For a recent comparison between these two models, see \citet{brightman15a}.

The BNTorus and MYTorus models measure similar line-of-sight column densities, $\nh= 9.92^{+0.13}_{-0.25}\times10^{22}$\,cm$^{-2}$ and $11.00^{+1.53}_{-0.20}\times10^{22}$\,cm$^{-2}$, respectively. The lower column densities compared to the previous models are due to the fact that the torus models also include Compton scattering, while \texttt{phabs} does not, which leads to an overestimation of the column in the latter. This is also reflected in the slightly lower unabsorbed 3--50\,keV luminosity of the BNTorus model of $\approx3.1\times10^{42}$\,erg\,s$^{-1}$.
The opening angle of the torus measured by BNTorus is $60.00^{+0.13}_{-2.97}$ degrees, which corresponds to a covering factor of 0.5. This covering factor compares well to other local AGN of similar luminosity, such as NGC~1068, NGC~1320, and IC~2560 \citep{balokovic14a, bauer15a, brightman15a}. 

For MYTorus, the inclination angle of the torus is derived to be $\geq76$ degrees. MYTorus has the added flexibility of decoupling the scattered and fluorescent line components from the transmitted component in order to test for scattering out of the line of sight. However, when allowing for such a decoupling we only find marginal improvement in terms of  $\chi^2$ and the inclination angle becomes completely unconstrained. In that case we can place  an upper limit of $1.15\times10^{23}$\,cm$^{-2}$  on the \nh of any material out of the line of sight, consistent with what is seen along the line of sight.

Using \suz XIS and GSO data, \citet{fukazawa11a} found a significant reflection fraction of the order of $R\approx0.2$. Their best-fit model includes two power-law components describing the AGN core emission and the jet contribution separately. When applying their model to the \xmm and \nustar data we cannot confirm such high reflection fractions but instead obtain similar upper limits on $R$ as in the simpler models presented in Table~\ref{tab:par_full}. Following \citet{fukazawa11a} and using the \texttt{pexmon} model to self-consistently describe the \feka line and fixing the photon-indices at 1.6 and 1.9, respectively, we obtain $R=0.138\pm0.016$. However, the fit is clearly worse than the fits with only a single power-law ($\redchi=1.16$ for 1531 d.o.f). 

We also investigated the presence of a partial covering model for the primary absorber, as used by, e.g., \citet{evans04a}. Because we only consider data above 3\,keV, our limits are only marginally constraining, and we find a covering fraction $> 0.98$. In Section~\ref{susec:diff} we extend the energy range down to 2\,keV and find weak evidence for partial covering.

Using \suz data, \citet{tombesi14a} found evidence for two weak absorption lines at 6.66\,keV and 6.95\,keV, which they interpreted as evidence for a slow wind. Similar absorption lines have recently be discovered in the \nustar spectrum of Cyg~A, a bright FRII galaxy \citep{reynolds15a}. When adding Gaussian absorption lines to our data of \cena, with the energies fixed at the values found by \citet{tombesi14a}  and the width set to 1\,eV, we find a marginal improvement of $\Delta \chi^2 = 7$ for two additional parameters. However, if we allow the energies to vary the fit does not converge. The failure to detect significant absorption features could be due to the much lower \snr in the \xmm data compared to the \suz data used by \citet{tombesi14a}. We therefore do not include these lines in our discussion.

\begin{deluxetable*}{rllllll}
\tablewidth{0pc}
\tablecaption{Model parameters for the simultaneous \textsl{NuSTAR} and \textsl{XMM} spectra.\label{tab:par_full}}
\tablehead{\colhead{Parameter} & \colhead{Powerlaw} & \colhead{Cutoff-PL} & \colhead{pexrav}  & \colhead{compPS} & \colhead{MYtorus} & \colhead{BNtorus}} 
\startdata
 $ N_\text{H}~[10^{22}\,\text{cm}^{-2}]$ & $17.06^{+0.26}_{-0.24}$ & $16.78\pm0.26$ & $16.79^{+0.26}_{-0.25}$ & $16.86^{+0.30}_{-0.31}$ & $11.00^{+1.53}_{-0.20}$ & $9.92^{+0.14}_{-0.25}$ \\
 $ \mathcal{F}_\text{cont}\tablenotemark{a}$ & $0.9946\pm0.0024$ & $0.9896\pm0.0024$ & $0.9918\pm0.0024$ & $0.9936^{+0.0028}_{-0.0024}$ & --- & --- \\
 $ \Gamma$ & $1.815\pm0.005$ & $1.797\pm0.005$ & $1.797\pm0.005$ & --- & $1.824\pm0.006$ & $1.826^{+0.009}_{-0.008}$ \\
 $ E_\text{fold} \text{~or~} kT~[\text{keV}]$ & --- & $\left(1.000^{+0.000}_{-0.075}\right)\times10^{3}$ & $\left(1.000^{+0.000}_{-0.054}\right)\times10^{3}$ & $\left(2.16^{+0.19}_{-0.22}\right)\times10^{2}$ & --- & --- \\
 $ R$ & --- & --- & $\le0.011$ & $\le0.012$ & --- & --- \\
 $ y$ & --- & --- & --- & $0.402\pm0.016$ & --- & --- \\
 $ i~[\text{deg}]$ & --- & --- & 60 (fix) & 60 (fix) & $>75.8$ & $63.30^{+4.29}_{-0.11}$ \\
 $ \Theta_\text{tor}~[\text{deg}]$ & --- & --- & --- & --- & 60 (fix) & $60.00^{+0.13}_{-2.97}$ \\
 $ I_\text{Fe}\tablenotemark{b}$ & $\left(2.76\pm0.22\right)\times10^{-4}$ & $\left(2.88\pm0.22\right)\times10^{-4}$ & $\left(2.86\pm0.22\right)\times10^{-4}$ & $\left(3.38\pm0.26\right)\times10^{-4}$ & --- & --- \\
 $ E_\text{Fe}~[\text{keV}]$ & $6.404^{+0.005}_{-0.009}$ & $6.404^{+0.004}_{-0.008}$ & $6.402^{+0.007}_{-0.006}$ & $6.404^{+0.004}_{-0.007}$ & --- & --- \\
 $ \sigma_\text{Fe}~[\text{eV}]$ & $\leq8.7$ & $\leq8.8$ & $\leq8.5$ & $\leq8.7$ & --- & --- \\
 $ CC_\text{FPMB}$ & $1.0366\pm0.0028$ & $1.0366\pm0.0028$ & $1.0366^{+0.0028}_{-0.0026}$ & $1.0366\pm0.0028$ & $1.032\pm0.004$ & $1.032\pm0.004$ \\
 $ CC_\text{pn}$ & $0.848\pm0.007$ & $0.847\pm0.007$ & $0.847\pm0.007$ & $0.847\pm0.007$ & $0.866\pm0.009$ & $0.869\pm0.009$ \\
 $ CC_\text{MOS1}$ & $1.214\pm0.016$ & $1.212\pm0.016$ & $1.212\pm0.016$ & $1.213^{+0.016}_{-0.014}$ & $1.109^{+0.019}_{-0.018}$ & $1.116^{+0.019}_{-0.018}$ \\
 $ CC_\text{MOS2}$ & $1.238\pm0.016$ & $1.236\pm0.016$ & $1.237\pm0.016$ & $1.237^{+0.016}_{-0.015}$ & $1.128\pm0.019$ & $1.135\pm0.019$ \\
\hline$\chi^2/\text{d.o.f.}$   & 1595.50/1532& 1620.63/1531& 1620.67/1530& 1595.72/1531& 1667.04/1536& 1695.77/1535\\$\chi^2_\text{red}$   & 1.041& 1.059& 1.059& 1.042& 1.085& 1.105\enddata
\tablenotetext{a}{unabsorbed flux in keV\,s$^{-1}$\,cm$^{-2}$ [3--50\,keV]}\tablenotetext{b}{in ph\,s$^{-1}$\,cm$^{-2}$}
\end{deluxetable*}


\subsection{Contribution from the diffuse emission}
\label{susec:diff}

In the preceding section we attributed differences between the \xmm and \nustar spectra to pile-up and cross-calibration differences. The strength of these effects required to explain the differences is, while not impossible, somewhat surprising. We therefore made an effort to rule out astrophysical or source intrinsic effects that could cause this discrepancy. The main source of intrinsic background contributing to the measured spectrum is  diffuse emission surrounding the AGN as seen with \chandra. Due to the different PSF sizes of \xmm and \nustar, the instruments sample different amounts of this diffuse emission, which might influence the observed spectral slope.

To check the influence of the diffuse emission as a function of distance to the AGN, we extracted spectra in different annuli from the \xmm cameras. For the pn camera we use rings with 5\asec--15\asec, 15\asec--25\asec, and 25\asec--40\asec. We chose to avoid the central 5\asec to ensure the innermost pixel is excluded given pn's pixel size of 4.1\asec. For MOS\,1 and 2 we use annuli with  15--20\asec, 20\asec--40\asec, 40\asec--60\asec, 60\asec--80\asec, and 80\asec--100\asec.

For \nustar we used an extraction region of 100\asec, as described in Section~\ref{susec:nustar}. We also extracted spectra from smaller regions (10\asec and 40\asec) but did not find a significant difference in the spectral shape. We therefore chose to use the largest region for the best \snr. 

 We then fitted all these spectra simultaneously using a absorbed power-law  plus a Gaussian iron line. We required that all data have the same absorption column, photon index, and iron line energy, i.e., only allowed for the normalization of the continuum and the line to be different between the data sets. The iron line width was fixed to $10^{-6}$\,keV, far below the energy resolution of any of the instruments.

\begin{figure}
\centering
\includegraphics[width=0.95\columnwidth]{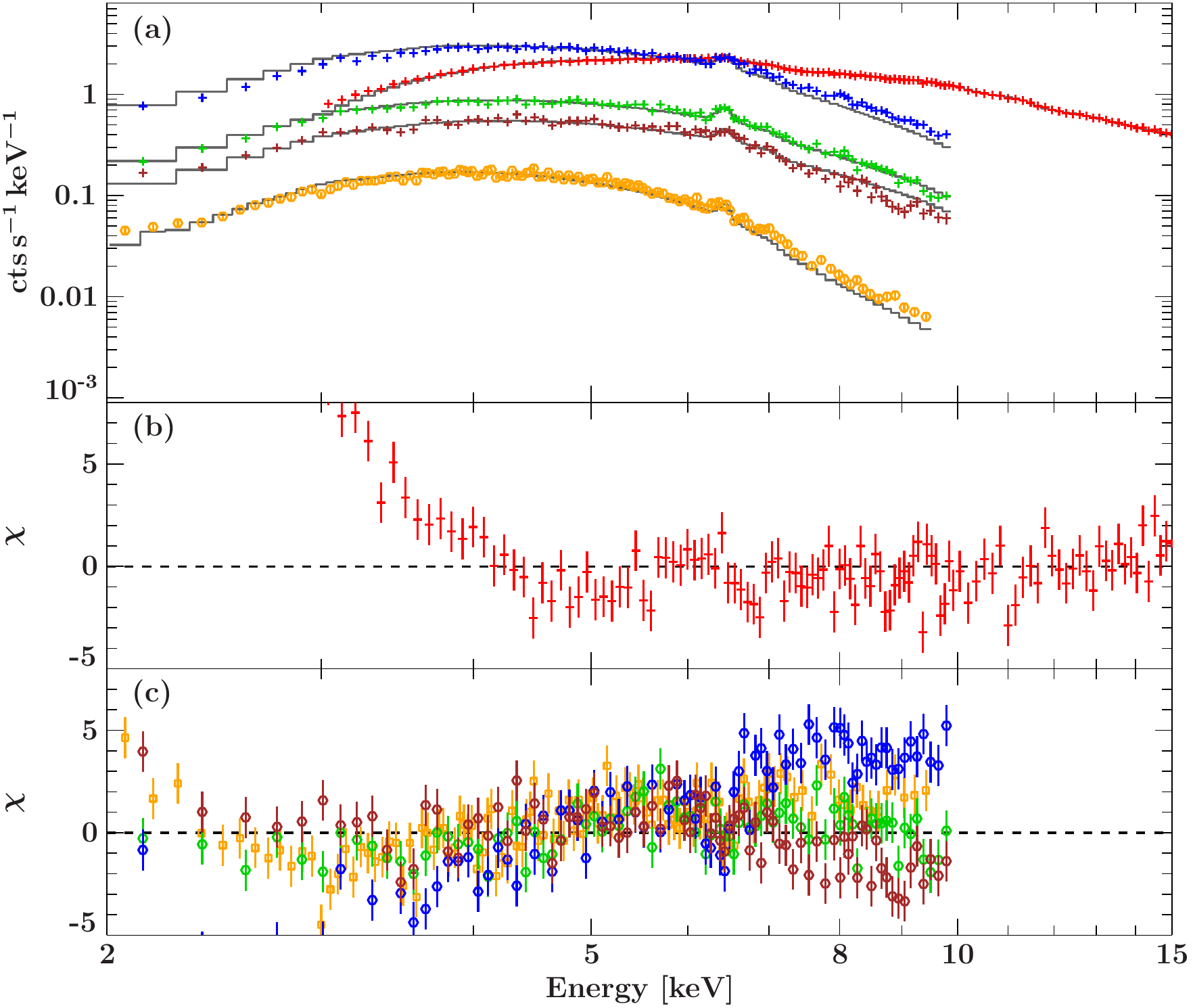}
\caption{\textit{(a)}: Spectra and best-fit models in different annuli using the measured background and a partially-absorbed power-law model. \nustar FPMA data are shown in red, \xmm pn data between 5--15\asec in blue, between 15--25\asec in green and between 25--40\asec in brown. For \xmm MOS\,1 residuals of all five annuli were combined into one spectrum for visual clarity, shown in orange. Data from \nustar/FPMB and \xmm/MOS\,2 are not shown for clarity. \textit{(b)} residuals in terms of $\chi$ for the \nustar data. \textit{(c)} residuals for the \xmm data.}
\label{fig:spec_rings_broadband}
\end{figure}

When restricting the energy range to 5--78\,keV for \nustar and 3--10\,keV for \xmm, we obtain a fit with similar values as the power-law fit in Table~\ref{tab:par_full} (model A, see Table~\ref{tab:par_rings} in the Appendix), but with a worse statistical quality ($\redchi=1.38$ for 1699 dof).  When  extending the energy range for \nustar down to 3\,keV and for \xmm down to 2\,keV we do not find a statistically acceptable fit, even when allowing for a partial covering absorber ($\redchi = 2.14$ for 1896 dof). 
Besides the clear mismatch of \nustar between 3--5\,keV,  the strongest residuals are due to the pn data, as shown in Fig.~\ref{fig:spec_rings_broadband}. 
These data show different spectral slopes for different extraction region sizes, which might indicate spatial variation of the spectrum due to diffuse emission. This diffuse emission might also influence the \nustar spectrum between 3--5\,keV and could be responsible for the observed discrepancies with \xmm. 

To investigate this we simulated how the diffuse emission, as seen with \chandra, influences the background in the different instruments and extraction regions. Details of the simulations are given in the Appendix. From these simulations it becomes clear that the diffuse emission cannot contribute enough flux to alter the observed \xmm and \nustar spectra significanlty. The core emission dominates over the diffuse background, even at large extraction radii. Even when allowing a scaling factor as a free parameter for each  background, we do not obtain a good fit ($\redchi=1.24$ for 1673 dof) and the scaling factors reach unrealistic values (e.g., almost 3 for pn, i.e., the pn background needs to be three times higher than measured with \chandra). 

 We conclude from this investigation that the observed discrepancies between \nustar and \xmm are attributed to pile-up and cross-calibration differences and that the diffuse emission around the AGN does not significantly influence the observed spectra. This result also implies that the measured data are completely dominated by the AGN itself and we obtain a clear view of the hard X-ray emission close to the central engine.

\section{Discussion}
\label{sec:disc}

We used simultaneous \xmm and \nustar data to measure the AGN emission of \cena with the best \snr yet and to study the origin of the hard X-ray emission. We find that a simple absorbed power-law or a thermal Comptonization spectrum with an \feka emission line fits the 3--78\,keV data very well. 
We do not find evidence for either an exponential rollover at high energies, a reflection component or a partial covering absorber, and put stringent upper limits on the folding energy and reflection fraction ($>1$\,MeV and $<0.01$ respectively). 

\subsection{The origin of  the iron line}

Many radio-loud AGN that are not pure blazar, have a narrow \feka line with no indication of reflection from a disk close to the black hole and only weak evidence for distant reflection (e.g., 3C~33, \citealt{evans10a}; 3C~382, \citealt{ballantyne14a}; and 3C~273, \citealt{madsen15a}, see also \citealt{wozniak98a}). 
The lack of relativistically blurred reflection has been discussed extensively in the literature, with the most common explanations being either an ionized inner accretion disk \citep{ballantyne02a}, a slightly truncated inner accretion disk due to retrograde spin \citep{garofalo09a}, or an outflowing corona \citep[although their model predicts a significantly higher reflection strength for the measured photon index of \cena]{malzac01a}. 
Weak and very weak reflection features are therefore not unusual in radio-loud AGN, like we find for the \nustar spectrum of \cena. 

The narrow \feka line likely originates from absorbing material relatively far away from the core. As shown by \citet{rivers11b}, the absorber in \cena is not Compton thick, but is thick enough to produce the observed \feka line strength. In fact, assuming that a spherically symmetric absorbing medium surrounding the X-ray source is responsible for the observed \feka emission, the predicted equivalent width is much higher than observed.
%
Following the calculations of \citet{markowitz07a}, for a measured column density of $\nh\approx1.7\times10^{23}$\,cm$^{-2}$ we obtain $\text{EW}_\text{calc}=109$\,eV, compared to $\approx 40$\,eV observed. As discussed by \citet{markowitz07a} a spherically symmetric shell is a very simplified geometry, and if the  absorber is only partially covering the X-ray source, the equivalent width will be reduced. Furthermore the calculation assumes solar abundances and the equivalent width can be significantly reduced with a sub-solar iron abundance.

A more realistic absorber geometry is a torus configuration, as invoked for many Compton-thick AGN and as suggested from the unification scheme \citep[see, e.g.][]{antonucci93a}. As demonstrated by 
\citet{matt03a}, column densities around $\nh\approx10^{23}$\,cm$^{-2}$ will lead to equivalent widths on the order of 40--50\,eV, while not producing any significant Compton hump. As we have shown, physically motivated torus models (MYTorus, BNTorus) describe the data very well and self-consistently explain the strength of the iron line.

Infrared photometry of \cena can also be well described with a (clumpy) torus model, with the caveat that the contribution of synchrotron emission to the IR data is not known \citep{ramosalmeida09a}. From these IR models a column density around $\nh=6.6^{+2.2}_{-1.8}\times10^{23}$\,cm$^{-2}$ for the torus is inferred, similar to the absorption column measured in the X-rays.

As shown by \citet{rothschild06a}, using \xte data taken between 1996--2009 and compared them to previous studies, the flux of the iron line is stable over long time scales ($>$10\,yr). 
We confirm these results, and measure $I_\text{Fe}=\left(2.76\pm0.22\right)\times10^{-4}$\,ph\,cm$^{-2}$\,s$^{-1}$. Similar values have been seen in \suz: ($\left[2.3\pm0.1\right]\times10^{-4}$\,ph\,cm$^{-2}$\,s$^{-1}$; \citealt{markowitz07a}, and [2.7--3.0]$\times10^{-4}$\,ph\,cm$^{-2}$\,s$^{-1}$; \citealt{fukazawa11a}), \sax \citep[$2.7^{+0.8}_{-1.4}\times10^{-4}$\,ph\,cm$^{-2}$\,s$^{-1}$;][]{grandi03a}, and \xmm \citep[$\approx2.4\times10^{-4}$\,ph\,cm$^{-2}$\,s$^{-1}$;][]{evans04a}. 

On the other hand, the continuum flux is strongly variable, by more than a factor of two \citep[e.g.,][]{rothschild06a}. The flux presented here is about 40\% higher than the average long-term flux observed by \inte \citep[averaged over 6 years between 2003--2009,][]{beckmann11a}. This results in a strong variability of the equivalent width of the iron line and limits the applicability of using the instantaneous X-ray flux to calculate the equivalent width.
To explain the stability of the \feka flux, the fluorescent region  needs to be on the order of 10\,ly or more away from the core, to smear out its variations on that time-scale. The region can still be much smaller than resolvable even with \chandra (as $1\asec$ is about 55\,ly at the distance of \cena).

\subsection{Spectral curvature at high energies}

Seyfert galaxies produce hard X-rays through thermal Comptonization of soft seed photons in a hot electron-gas corona. The temperature of the corona can be estimated from the energy of the exponential rollover, however, care has to be taken since the \texttt{cutoffpl} model has a distinctly different shape than calculations of a Comptonization spectrum \citep[see, e.g.,][]{petrucci01a}.
 \nustar has measured folding energies in numerous Seyfert galaxies, e.g., IC 4329A \citep[$186\pm14$\,keV; ][]{brenneman14a}, SWIFT~J2127.4+5654 \citep[$108 \pm 11$\,keV; ][]{marinucci14a}, MCG$-$05-23-016 \citep[$116\pm6$\,keV;][]{balokovic15a}, as well as determined lower limits in NGC 5506 with $>350$\,keV and a best fit $\approx720$\,keV \citep{matt15a}. \citet{fabian15a} summarize and discuss these measurements. 
 Recently, \nustar observations of the broad-line radio galaxy 3C~390.3 revealed a folding energy of $117^{+18}_{-14}$\,keV \citep{lohfink15a}, much lower than we find for \cena. 
 In \cena the lower limit  is in excess of 1\,MeV, which, if the continuum is produced in a thermal corona, indicates a very high plasma temperature.
 
 Following the calculations by \citet{fabian15a} this very high temperature would put \cena's corona far above the pair-production line for a coronal size of 10\,$r_g$. Only a corona orders of magnitude larger than typically measured for other AGN would place \cena in the physically allowed regime. 
However, the phenomenological nature of the \texttt{cutoffpl} model makes a physical interpretation difficult. 
A more realistic estimate of the temperature can be obtained using the thermal Comptonization  \texttt{compps} model, which gives $kT_e=216^{+19}_{-22}$\,keV assuming a slab geometry and a seed photon tempeature $kT_\texttt{BB}=10$\,eV. 
This temperature is stable against different geometries but depends on the seed photon temperature and spectral distribution. We find $kT_e$ to be between 100--300\,keV for input temperatures between 5--50\,eV.
Our results are consistent with the one measured by \inte for $kT_\texttt{BB}=10$\,eV but statistically better constrained \citep[$kT_e=206\pm62$\,keV,][]{beckmann11a} and, assuming a slightly extended corona of $\sim100\,r_g$, are in line with the pair-production limit.
 
The value of the folding energy  of \cena is discussed extensively in the literature, with no clear consensus. 
For example, \citet{rothschild06a} measure a folding energy $>1.5$\,MeV using \xte while at a similar luminosity \citet{kinzer95a} find $E_\text{cut} = 254\pm33$\,keV using \gro/OSSE data. 
From the fluxes and spectral shape measured between 0.2--30\,GeV with \fermi it is clear that the spectrum needs to rollover or break somewhere in the 100-1000\,keV range \citep{fermicena10a}.

It is interesting to note that nearly all well constrained measurements of a folding energy were performed by $\gamma$-ray instruments sensitive at energies $>100$\,keV, while purely X-ray missions often find very high lower limits of the folding energy far outside their covered energy range. As discussed above, this effect is likely connected to the difference between a \texttt{cutoffpl} and a realistic Comptonization model: the \texttt{cutoffpl} model is constantly curving, even far below the folding energy, while a realistic Compton spectrum is much more power law-like at energies significantly below the temperature of the Comptonization plasma and rolls over  more steeply than the \texttt{cutoffpl} above it \citep[see Figure~3 in][and references therein]{fabian15a}.
The $\gamma$-ray instruments like \inte therefore detect the cutoff, but given their typically lower statistics at soft X-rays find an acceptable solution with a \texttt{cutoffpl} or a broken power-law model \citep{kinzer95a, beckmann11a}. For the X-ray instruments, on the other hand, the rollover is outside their energy range and they mainly measure the power-law part of the Comptonization spectrum, resulting in unconstrained or very high folding energies when using \texttt{cutoffpl}. By using a more physical Compton spectrum  we obtain a statistically well constrained measurement and show that a temperature between 100--300\,keV is in line with the observed spectra.
We note that the seed photon spectrum in an ADAF flow is not necessarily described by a multi-temperature black-body spectrum. However, by sampling of a wide range of input temperature we demonstrate that the measured cutoff depends only weakly on the exact seed photon spectrum.

\subsection{The geometry and physics of the X-ray corona}

Despite the exceptional quality of the \xmm and \nustar data, the origin of the hard X-rays cannot be uniquely determined. 
Both models are consistent with the broad-band spectral energy distribution (SED) presented by \citet{fermicena10a}. To better constrain which emission mechanism is dominant in \cena modeling a simultaneous SED is necessary which will be presented in a forthcoming work (M\"uller et al., inprep). 
We rule out any contribution from reflection from the inner accretion disk  with high significance, similar to the X-ray spectra of other radio galaxies.
This measurement is in line with the idea that
the hard X-ray emission from \cena is dominated by SSC emission from the inner radio jet \citep{mushotzky78a,fermicena10a}. In this model the X-rays are produced in an outflowing plasma by Compton up-scattering synchrotron seed photons, and it explains well the broad-band SED other than the TeV $\gamma$-ray flux detected by H.E.S.S. \citep{fermicena10a, hesscena09a}.

\citet{beckmann11a} remark, however, that a jet origin of the hard X-rays is more difficult to reconcile with the small long-term variability of the X-ray flux, which is more reminiscent of Seyfert galaxies. 
A possible solution includes contribution from both components, a thermal corona as well as a synchrotron jet \citep{soldi14a}. Such a combined model has been proposed for other radio galaxies as well, such as 3C~120 \citep{lohfink14a} and  3C~273 \citep{grandi04a,madsen15a}. However, as \citet{rothschild06a} and later \citet{burke14a} found, the X-ray continuum shape is remarkably stable over time, despite significant flux changes. If the flux variability were induced by the inner jet component, we would expect some influence on the hard X-ray continuum. On the other hand, variability of the cutoff-energy as a function of flux has been observed with soft $\gamma$-ray instruments \citep[e.g., with \gro, ][]{kinzer95a}, following the ``softer-when-brighter'' correlation of Seyfert galaxies.

Some authors have reported a significant reflection fraction in \cena \citep[e.g.,][]{fukazawa11a, burke14a}. If these detections are real, they do not seem to correlate with a particularly weak state of the X-ray flux, which we would expect if high fluxes correspond to a strong contribution from the jet emission, smearing out the reflection component. In particular, the \inte/SPI data used by \citet{burke14a} are an average over 10 years, while  \citet{fukazawa11a} report a similar reflection fraction in both low and high flux states corresponding to a flux change of almost a factor of 2. 
A mixture of standard thermal Comptonization and jet emission, in which the jet is driving the observed variability, thus seems unlikely.

If a stable accretion disk is present, we need to obscure it completely to eliminate all evidence of reflection from the observed spectrum. A puffed up accretion disk with a small corona could result in such an observed spectrum. However, \cena is only accreting at $<0.2\%$ of its Eddington luminosity, making a geometrically thick accretion disk unlikely \citep{paltani98a}. Rather, the accretion disk might be strongly truncated and replaced with an optically thin accretion flow, as in the advection-dominated accretion flow (ADAF) model \citep{narayan95a}.

\citet{rieger09a} propose that \cena is dominated by ADAF emission, which they use to predict that \cena might be a source of TeV photons and ultra-high energy (UHE) cosmic-rays. While the latter claim is disputed in the literature \citep[who instead favor a two-zone SSC model, with UHE particles emerging from the lobes, but see also \citealt{khiali15a}, for a model using magnetic reconnection to produce $\gamma$-rays]{petropoulou14a}, a large ADAF can explain the observed hard X-ray properties. Typical temperatures for the electrons in an ADAF Comptonization plasma are on the order of 
100\,keV, in good agreement with our measurement.

The fact that the \nustar spectrum is rather simple and well described by one power-law or Comptonization component also argues against a mix of X-ray sources and would instead seem to favor  a common origin for all observed hard X-rays.

\section{Summary and Outlook}
\label{sec:summ}

Using the exceptional quality of simultaneous \nustar and \xmm spectra, we find that the core spectrum of \cena can be described by  a simple absorbed power-law with a photon-index $\Gamma\approx1.8$ or a single-temperature Comptonization spectrum. The phenomenological \texttt{cutoffpl} does not provide a good fit and we argue that this is due to the fact that its shape does not represent a realistic Comptonization spectrum. From the Comptonization model, we find a  coronal temperature of $kT_e\approx220$\,keV, for an assumed seed photon temperature of 10\,eV.

We carefully analyzed the diffuse emission observed by \chandra, including the hot ISM, the outer jet, and off-nuclear point-sources,  and find that it does not significantly contribute to the observed hard X-ray spectrum from the core. The morphological and spectral analysis of the diffuse emission will be presented in a forthcoming publication (Graefe et al., in prep.). 

We put stringent upper limits on the contribution of Compton-thick reflection, with a reflection fraction $R<0.01$. This rules out a standard Seyfert-like production of the hard X-rays and indicates that the inner accretion disk is replaced by optically thin gas. Despite the lack of reflection, the prominent iron line  can be self-consistently described using a torus model, and we find inclinations marginally consistent with  the torus being perpendicular to the jet-axis. We argue that Comptonization in an ADAF flow or  at the base of the inner jet or both can explain the observed spectrum. Multi-epoch, multi-wavelength observations will help to disentangle the contribution from the jet and the ADAF and will be presented in a forthcoming publication (M\"uller et al., in prep.).

\acknowledgments
We thank the anonymous referee for their comments which helped to improve this work.
This work was supported under NASA Contract No. NNG08FD60C, and
made use of data from the {\it NuSTAR} mission, a project led by
the California Institute of Technology, managed by the Jet Propulsion
Laboratory, and funded by the National Aeronautics and Space
Administration. We thank the {\it NuSTAR} Operations, Software and
Calibration teams for support with the execution and analysis of these observations. This research has made use of the {\it NuSTAR}
Data Analysis Software (NuSTARDAS) jointly developed by the ASI
Science Data Center (ASDC, Italy) and the California Institute of
Technology (USA). 
Based on observations obtained with \xmm, an ESA science mission with instruments and contributions directly funded by ESA Member States and NASA.
This research has made use of a collection of ISIS scripts provided by the Dr. Karl Remeis observatory, Bamberg, Germany at \url{http://www.sternwarte.uni-erlangen.de/isis/}.
CM acknowledges the support of the Bundesministerium f\"ur Wirtschaft
und Technologie (BMWi) through Deutsches Zentrum f\"ur Luft- und
Raumfahrt (DLR) grant 50OR1404.
We acknowledge support of the Deutsche Forschungsgemeinschaft (DFG) through grant Wi 1860 10-1.
This research was funded in part by NASA through Fermi Guest
Investigator grants NNH09ZDA001N, NNH10ZDA001N, NNH12ZDA001N,
NNH13ZDA001N-FERMI. This research was supported by an appointment to
the NASA Postdoctoral Program at the Goddard Space Flight Center,
administered by Oak Ridge Associated Universities through a contract
with NASA. MB acknowledges support from NASA Headquarters under the NASA Earth and Space Science Fellowship Program, grant NNX14AQ07H.
We would like to thank John E. Davis for the \texttt{slxfig} module, which was used to produce all figures in this work. 

{\it Facilities:} \facility{NuSTAR}, \facility{XMM}, \facility{CXO}

\begin{appendix}

\subsection{The diffuse emission as seen with \chandra}
\label{susec:chandra_diff}
\cena is known to show a complex morphology in X-rays as seen with \textsl{Einstein}, \textsl{ROSAT}, and \chandra \citep{feigelson81a,turner97a, kraft00a, kraft02a, evans04a}. In addition to an extended X-ray jet and point-sources in the host galaxy, the AGN is surrounded by faint diffuse emission, extending about $1'$ ($\approx$1\,kpc) around the core. While this diffuse emission is not visible in the MOS image (Fig.~\ref{fig:image}, right),  it still might contribute to the observed X-ray spectrum. We therefore need to find a model for the extended emission, which can be added to the modeling of the \xmm and \nustar data. Such  a model can only be obtained from \chandra, due to its higher angular resolution. A detailed discussion of the \chandra data will be presented in a forthcoming publication (Graefe et al., in prep.), while here we only concentrate on its influence to the background.

\cena has been observed multiple times by \chandra with both sets of CCDs of the Advanced CCD Imaging Spectrometer \citep[ACIS;][]{chandraref}, but not simultaneous with \nustar and \xmm. We selected a 5.34\,ks ACIS-I observation (ObsID 15295; PI S. Murray) taken on 31 August 2013 which is in closest proximity to our \nustar observation.
This observation, however, was too short to provide sufficient \snr to describe the diffuse spectrum accurately. We therefore looked through the archive for observations at a similar flux level and similar spectral shape and selected four of the longest ACIS-I exposures taken in 2007 (ObsID 7797-7800, PI R.~Kraft; see Table~\ref{tab:obsdates} for an overview of the data used.). 
 We reprocessed each observation using {\sc ciao} version 4.5 to create new level 2 event files, following the software threads from the \chandra X-ray Center (CXC)\footnote{\url{http://cxc.harvard.edu/ciao}}. 

We used the {\sc specextract} task to extract X-ray spectra in each event file for several annular apertures centered on ($13^{h}25^{m}27.59^{s}$, $-43^{d}01^{m}08.95^{s}$) with radii of 5--15\asec, 15--25\asec, 25--40\asec, 40--100\asec, and 8--100\asec, i.e. matching the pn annuli.  An exclusion aperture 3$''$ wide was placed on each read-out streak, the direction of which varied with each exposure.  The inner 5\asec were too piled-up for spectral extraction.  A background spectrum was simultaneously extracted from the same chip in a sourceless region, and automatically scaled based on the ratio of the source-to-background areas. We then combined the spectra from the four longest exposures for each annulus, using the {\sc combine\_spectra} task, which also calculates the combined background spectrum and response files. 
The spectra of the shallow, recent spectrum (ObsID 15295) shows little variation with respect to the deep, combined spectrum so that we base our analysis on the combined March 2007 data.

We fitted a partially covered power-law to the data, requiring that all annuli have the same absorption column and covering fraction, but allowed for different photon indices and normalizations. This model is purely phenomenological and allows to account for diffuse emission leaking at the lowest energies. We additionally added a narrow \feka line around 6.4\,keV.  The best-fit parameters for this model are given in Table~\ref{tab:par_chanonly}. This model resulted in a very good fit, with \redchi=1.05 for 1630 dof. Adding an exponential rollover to the model by replacing the power-law with the XSPEC \texttt{cutoffpl} model did not improve the fit and resulted in an unconstrained folding energy.

\begin{figure}
\centering
\includegraphics[width=0.45\columnwidth]{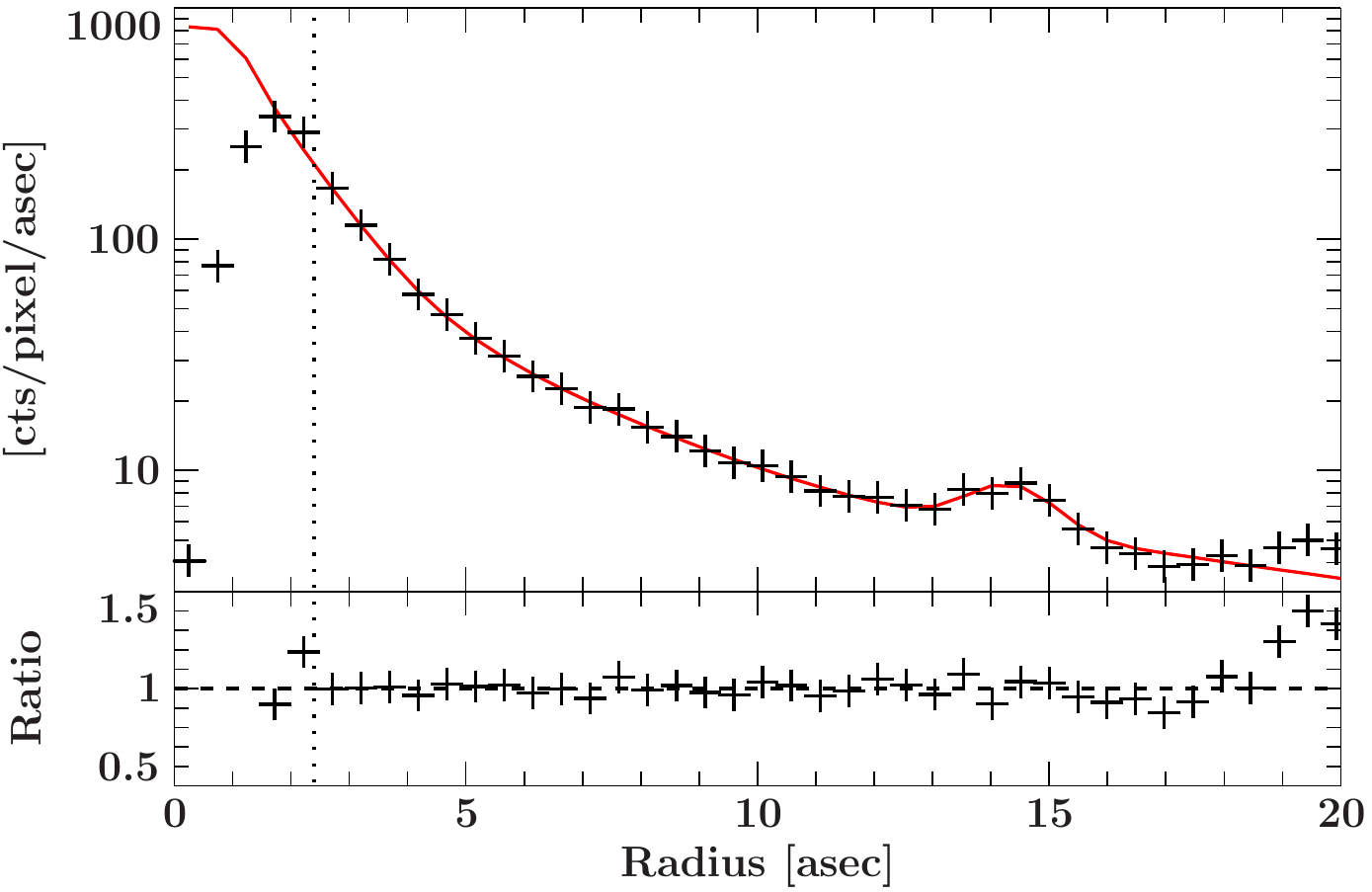}
\caption{\textit{(a)} Radial profile of the \chandra image of \cena, binned into 1\,pixel size bins (0.492$''$) as function of distance from the core. The dotted line indicates 2.5$''$, below which the data were ignored for the fit due to pile-up. The best fit model is shown in red. The excesses at $\approx14''$ and $\approx20''$ are due to point-sources. \textit{b)} Data-to-model ratio of the best fit.}
\label{fig:chandraprofile}
\end{figure}

The core of \cena is so strongly piled-up that no events are registered at the center. Pile-up continues to be high out to $\approx2.5''$. However, any diffuse emission in that region will also contribute to the diffuse background in \xmm in the annuli outside of $3''$, as the PSF of \xmm has a FWHM of about $4.5''$. We therefore try to estimate the contribution of the diffuse emission under the core by extrapolating the density profile of the \chandra image (using data from observation 7797 only). To do that, we construct the radial intensity profile centered at the core of \cena by binning the events in a linear grid with one pixel (0.492$''$) spacing as function of distance from the core. The profile is shown in Figure~\ref{fig:chandraprofile}. The intensity drops dramatically inward of  $2.5''$ due to the very high pile-up.

We describe this profile between 2.4--20$''$ with a broken power-law plus a zero-centered Gaussian to estimate the contribution within the center. We set the break value of the broken power-law to $1''$ and the power-law index below that break to 0, to prevent the power-law from rising to infinity at the center. Instead the center is described by a Gaussian function with a width of $\sigma=1.63''$. Using a $\beta$-model  \citep{kraft03a} instead of a power-law does not change the result, as at the relevant distances from the core the power-law part of the $\beta$-model dominates. 
We add another Gaussian line around 14.3$''$ to describe the excess produced by a weak source. As can be seen in Figure~\ref{fig:chandraprofile}, this model describes the radial profile very well. The exact rate of the center is not well constrained and we estimate  our systematic uncertainties  to be around a factor of 1.5--2.

\begin{figure}
\centering
\includegraphics[width=0.95\columnwidth]{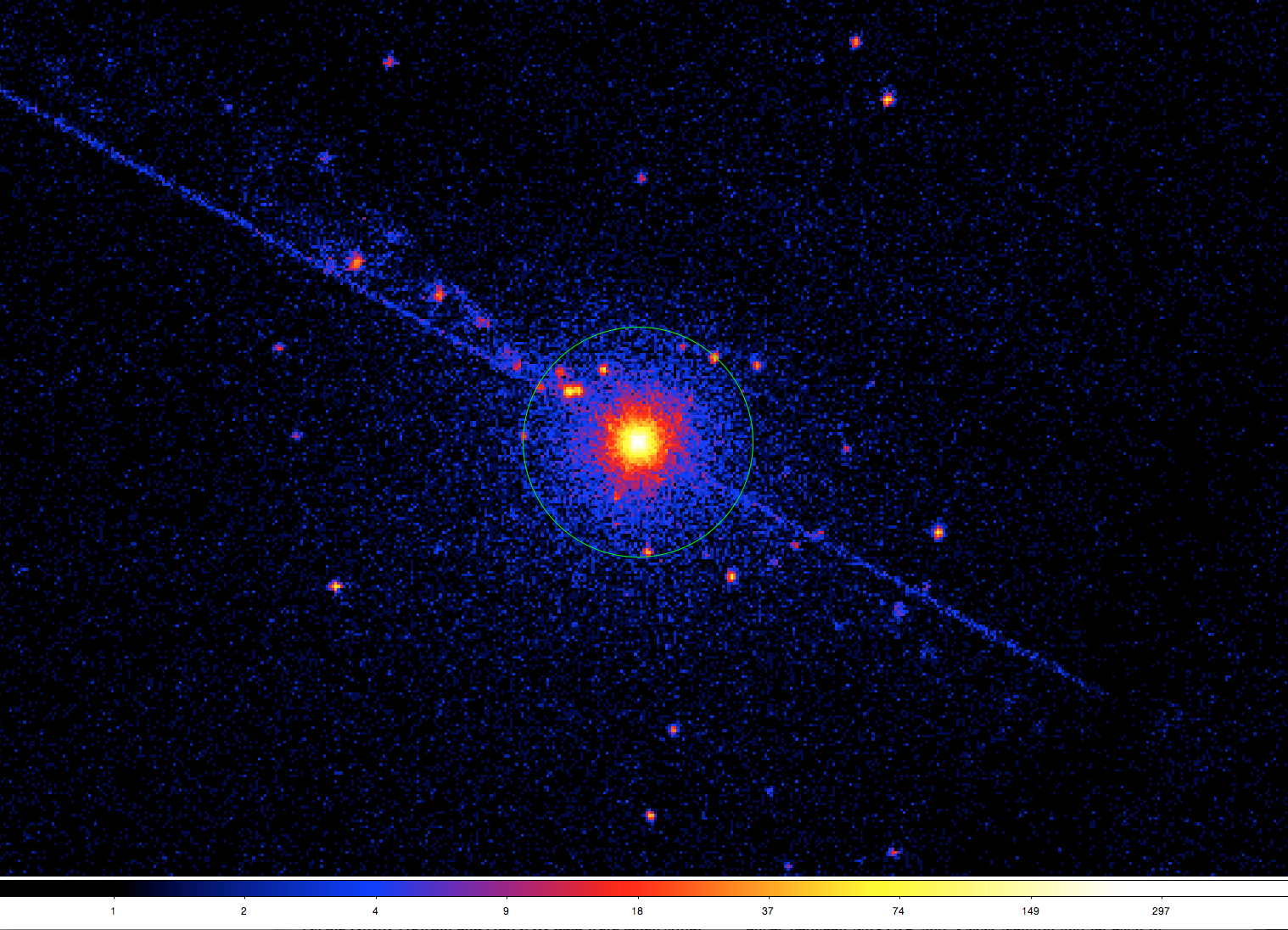}
\caption{\chandra image of the \cena core after filling in the piled-up core with or best estimate for the diffuse emission. North is up, East is to the left. The green circle is $20\asec$ in radius.}
\label{fig:chandra_filled}
\end{figure}

Using this profile we fill in the piled-up region of the \chandra image, replacing the inner $5''$ with counts drawn from a Poisson statistic as predicted by our model. This results in a very smooth image, shown in Figure~\ref{fig:chandra_filled}. We use this spectrum as an input in simulating the contribution of the diffuse background in \xmm and \nustar.

\subsubsection{Diffuse emission simulation}

The modified \chandra image shown in Figure~\ref{fig:chandra_filled} was used as input to the simulations and convolved with the respective \nustar and \xmm PSF. The extracted annuli were defined as regions with constant spectral properties, and each region was simulated into a separate image. When setting extraction regions for \nustar and \xmm, we calculated the relative contributions of each image (or spectra), and folded the weighted input spectra through the response files and then combined them into the output spectrum for the requested region. These simulated spectra were then used as new background spectra for the  \nustar and \xmm data.

Because of the relatively modest extent of \cena ($\approx$100\asec) and the scale of the extraction regions ($\approx$20--40\asec), we made the following approximations to simplify the simulations: we assumed a flat effective area coinciding with the center of the object rather than a continuous extended effective area of the underlying diffuse component. This approximation is valid since most of the emission originates in the inner few arc-seconds, dominating the response, and because at small off-axis angles ($<2'$) the extended effective area of a circular region cancels out to the area obtained from the center of a circle. In addition we did not include an energy-dependent PSF, since the effect is typically on the order of a few arc-seconds, while the scale size of our simulations was probing changes on tens of arc-seconds scale.  

\subsubsection{Results}

We use the emission as estimated from the \chandra data as background for the different \xmm annuli and the \nustar spectrum. We then fit the \xmm and \nustar data between 2--10\,keV and  3--78\,keV, respectively, with a partially covered power-law and an iron line simultaneously, allowing for the normalization of the continuum and the \feka line to change.

This additional background changes the fit parameters significantly (e.g., the photon-index softens from $\Gamma=1.82$ to $\Gamma \approx 1.95$), as the diffuse spectrum is very hard and we have no handle on a possible cutoff outside of \chandra's energy range. This fit is statistically similar,  with $\redchi=2.13$ for 1739 dof. 
The reduced number of degrees of freedom is due to our binning to a certain \snr level, which requires stronger binning for the now higher background. Allowing for a high-energy cutoff by replacing the power-law with the \texttt{cutoffpl} model did not improve the fit significantly ($\redchi=2.02$ for 1738 dof) and gives a folding energy around $E_\text{fold}\approx150$\,keV. 

A better fit can be achieved by allowing the normalization of the background to vary (model \texttt{recorn} in XSPEC), individually for each \xmm and \nustar spectrum (while requiring FPMA and FPMB and each annulus of MOS\,1 and 2 to have the same scaling factor). This approach significantly improved the fit to $\redchi=1.62$ for 1730 dof.  
However, strong residuals in the \nustar data below 5\,keV are still present. We therefore rule out a significant contribution from the diffuse emission to the low-energy end of the \nustar spectrum.

By ignoring all \nustar data below 5\,keV and allowing for a free scaling of the background we obtain a very good fit with $\redchi=1.18$ for 1680 dof (model B). 
However,  the scaling factors are very widely spread with $CB_\text{FPMA}=0.2$ and $CB_\text{pn1}=2.5$, where pn1 denotes the factor for the innermost pn annulus between 5--15\asec. We give the best-fit parameters in Table~\ref{tab:par_rings}.

\begin{deluxetable}{rlll}
\tablewidth{0pc}
\tablecaption{Model parameters using simultaneous fits of different annuli in \textsl{XMM}.\label{tab:par_rings}}
\tablehead{\colhead{Parameter} & \colhead{Model A\tablenotemark{a}} & \colhead{Model B\tablenotemark{b}} & \colhead{Model C\tablenotemark{c}}} 
\startdata
 $ N_\text{H}~[10^{22}\,\text{cm}^{-2}]$ & $17.63\pm0.22$ & $16.9\pm0.4$ & $17.8\pm0.4$ \\
 $ A_\text{cont}\tablenotemark{a}$ & $0.2440\pm0.0027$ & $0.243\pm0.006$ & $0.253\pm0.007$ \\
 $ \text{CF}$ & --- & $0.9932^{+0.0022}_{-0.0020}$ & $0.9917^{+0.0021}_{-0.0020}$ \\
 $ \Gamma$ & $1.820^{+0.005}_{-0.004}$ & $1.831\pm0.014$ & $1.852\pm0.015$ \\
 $ E_\text{fold}~[\text{keV}]$ & --- & --- & $\left(1.29^{+0.17}_{-0.14}\right)\times10^{2}$ \\
 $ A_\text{Fe}\tablenotemark{a}$ & $\left(2.4\pm0.4\right)\times10^{-4}$ & $\left(2.4\pm0.4\right)\times10^{-4}$ & $\left(1.8\pm0.4\right)\times10^{-4}$ \\
 $ E_\text{Fe}~[\text{keV}]$ & $6.4500^{+0.0016}_{-0.0185}$ & $6.408^{+0.004}_{-0.007}$ & $6.407^{+0.005}_{-0.006}$ \\
 $ B_\text{FPM}$ & --- & $0.20\pm0.09$ & 1 (fix) \\
 $ B_\text{pn} (5$--$15'')$ & --- & $2.75\pm0.29$ & $2.76^{+0.28}_{-0.30}$ \\
 $ B_\text{pn} (15$--$25'')$ & --- & $0.5\pm0.4$ & $1.36^{+0.29}_{-0.30}$ \\
 $ B_\text{pn} (25$--$40'')$ & --- & $0.10^{+0.12}_{-0.00}$ & $0.60\pm0.26$ \\
 $ B_\text{MOS} (15$--$20'')$ & --- & $0.44\pm0.22$ & $0.48\pm0.22$ \\
 $ B_\text{MOS} (20$--$40'')$ & --- & $0.93\pm0.29$ & $1.02\pm0.29$ \\
 $ B_\text{MOS} (40$--$60'')$ & --- & $0.48\pm0.20$ & $0.53\pm0.20$ \\
 $ B_\text{MOS} (60$--$80'')$ & --- & $0.87\pm0.30$ & $0.91\pm0.30$ \\
 $ B_\text{MOS} (80$--$100'')$ & --- & $1.0^{+9.0}_{-0.9}$ & $1.0^{+9.0}_{-0.9}$ \\
\hline$\chi^2/\text{d.o.f.}$   & 2356.72/1699& 1989.06/1680& 2081.68/1673\\$\chi^2_\text{red}$   & 1.387& 1.184& 1.244\enddata
\tablenotetext{a}{Model A: power-law with measured background and \textsl{NuSTAR} between 5--79\,keV and \textsl{XMM} between 3--10\,keV.}\tablenotetext{b}{Model B: power-law with additional diffuse background with free background scaling factor for all spectra. \textsl{NuSTAR} between 5--79\,keV and \textsl{XMM} between 2--10\,keV.}\tablenotetext{c}{Model C: cutoff power-law with additional diffuse background where the background scaling factor for \textsl{NuSTAR}/FPMA is fixed at 1. \textsl{NuSTAR} between 5--79\,keV and \textsl{XMM} between 2--10\,keV.}
\end{deluxetable}

When forcing the scaling factor for \nustar to be 1, i.e., assuming that our simulations capture exactly the correct background, we only find an acceptable fit when at the same time allowing for an exponential high-energy rollover (using the \texttt{cutoffpl} model in XSPEC). This model gives $\redchi=1.25$ for 1675 dof (model C). The best-fit parameters are shown in Table~\ref{tab:par_rings}. Still the scaling factors for the background of the other instruments vary wildly, indicating that the diffuse emission is not driving the observed differences.

\begin{deluxetable*}{cccccccc}
\tablewidth{0pc}
\tablecaption{Model parameters for the \textsl{Chandra} annuli fits.\label{tab:par_chanonly}}
\tablehead{\colhead{Instrument} & \colhead{$N_\text{H}~[10^{22}\,\text{cm}^{-2}]$} & \colhead{$\text{CF}$} & \colhead{$\Gamma$} & \colhead{$E_\text{Fe}~[\text{keV}]$} & \colhead{$I_\text{cont}\tablenotemark{a}$} & \colhead{$I_\text{Fe}\tablenotemark{b}$} } 
\startdata
ACIS 5-15   & $25.6^{+1.9}_{-2.0}$  & $0.841^{+0.010}_{-0.011}$  & $0.76\pm0.09$  & $6.395^{+0.015}_{-0.016}$  & $\left(5.9^{+1.1}_{-1.0}\right)\times10^{-4}$  & $\left(1.21\pm0.18\right)\times10^{-5}$  \\
ACIS 15--25   & ---  & ---  & $1.09\pm0.10$  & $6.429^{+0.021}_{-0.030}$  & $\left(5.9^{+1.2}_{-1.0}\right)\times10^{-4}$  & $\left(6.1\pm1.3\right)\times10^{-6}$  \\
ACIS 25--40   & ---  & ---  & $1.13\pm0.10$  & $6.385^{+0.026}_{-0.025}$  & $\left(5.7^{+1.2}_{-1.0}\right)\times10^{-4}$  & $\left(6.5\pm1.3\right)\times10^{-6}$  \\
ACIS 40--100   & ---  & ---  & $1.61\pm0.10$  & $6.409^{+0.021}_{-0.030}$  & $\left(2.8^{+0.6}_{-0.5}\right)\times10^{-3}$  & $\left(9.9\pm2.1\right)\times10^{-6}$  \\
\enddata
\tablenotetext{a}{in ph\,s$^{-1}$\,cm$^{-2}$\,keV$^{-1}$ at 1\,keV.}\tablenotetext{b}{in ph\,s$^{-1}$\,cm$^{-2}$.}
\end{deluxetable*}

\end{appendix}

\end{document}